\documentclass[twocolumn,preprintnumbers,amsmath,amssymb,aps]{revtex4-1} 
\usepackage{amsmath} 
\usepackage[utf8]{inputenc} 
\usepackage[english]{babel} 
\usepackage{graphicx} 
\usepackage{listings} 
\usepackage{soul}
\usepackage{color}

\usepackage{multirow}

\usepackage{pifont}
\newcommand{\cmark}{\ding{51}}%
\newcommand{\xmark}{\ding{55}}%

\begin{document}

\title{On the effect of the thermostat in non-equilibrium molecular dynamics simulations}

\author{Jos\'e Ruiz-Franco$^{1}$, Lorenzo Rovigatti$^{1,2}$, Emanuela Zaccarelli$^{1,2}$}
\address{$^1$Dipartimento Di Fisica, Sapienza Universit\`a di Roma, P.le A. Moro 5, 00185 Roma, Italy}
\address{ $^2$CNR Institute of Complex Systems, Uos Sapienza, Roma, Italy}
\date{\today}

\begin{abstract}

The numerical investigation of the statics and dynamics of systems in nonequilibrium in general, and under shear flow in particular, has become more and more common. However, not all the numerical methods developed to simulate equilibrium systems can be successfully adapted to out-of-equilibrium cases. This is especially true for thermostats. Indeed, even though thermostats developed to work under equilibrium conditions sometimes display good agreement with rheology experiments, their performance rapidly degrades beyond weak dissipation and small shear rates.
Here we focus on gauging the relative performances of three thermostats, Langevin, dissipative particle dynamics, and Bussi-Donadio-Parrinello under varying parameters and external conditions. We compare their effectiveness by looking at different observables and clearly demonstrate that choosing the right thermostat (and its parameters) requires a careful evaluation of, at least, temperature, density and velocity profiles. We also show that small modifications of the Langevin and DPD thermostats greatly enhance their performance in a wide range of parameters.

\end{abstract}

\maketitle

\section{Introduction}

In molecular dynamics (MD) simulations, the motion of $N$ individual particles in a specified volume $V$ evolves according to Newton's laws. Hence, since the energy is conserved, from a thermodynamic point of view the resulting system is a microcanonical ensemble ($NVE$)~\cite{allen2017computer,frenkel2001understanding}. However, in order to mimic experimental conditions it is often necessary to simulate systems at constant temperature rather than energy, obtaining a canonical ensemble ($NVT$). The control of the temperature is achieved by coupling the system to a so-called ``thermostat'', which acts as a thermal bath. During the course of the years many different thermostats have been developed to not only reduce the side effects due to the coupling, but also to more accurately reproduce the phenomena observed in experiments. For instance, many thermostats that exhibit a good temperature control do not correctly reproduce hydrodynamics, which require local momentum conservation and Galilean invariance.
There exist methods that explicitly incorporate in the simulation solvent particles, such as multi-particle collision dynamics (MPCD)~\cite{ripoll2005dynamic,kapral2008multiparticle,gompper2009multi}, thus naturally building the correct hydrodynamics in the systems. However, these methods lie outside the scope of the present work, which deals with implicit solvent treatment only.

In the following \textit{excursus} we provide a non-comprehensive list of well-known thermostats. We consider only thermostats that take into account the effects of the solvent in an implicit way. We start with the well-known Berendsen thermostat \cite{berendsen1984molecular}. It consists in rescaling all particle velocities after a certain number of time steps so that the (instantaneous) kinetic energy matches the target one. This guarantees a constant thermodynamic temperature. The main drawback is that it does not sample the $NVT$ ensemble~\cite{bussi2007canonical}, which implies that it is dangerous to use it for data production. Furthermore, it is not Galilean invariant and it does not locally conserve momentum. 

A very simple, yet effective thermostat is the Andersen thermostat~\cite{andersen1980molecular}. Here the coupling between the system and a ``heat bath'' is explicit: particles undergo random collisions with the (fictitious) solvent, effectively acquiring new momenta extracted from a Maxwell-Boltzmann distribution corresponding to the desired temperature $T$. This thermostat is local, but it is not Galilean invariant. 

The Nose-Hoover (NH) thermostat comes next~\cite{nose1984unified,hoover1985canonical}. It introduces in the Hamiltonian of the system an additional internal degree of freedom which acts as an effective friction parameter and represents the thermostat coupling. Once again, this thermostat is non-Galilean invariant. Furthermore, the additional degree of freedom alters the dynamics, leading to artificial hydrodynamics~\cite{stoyanov2005molecular}. A common improvement over the original implementation is to consider Nose-Hoover chains, that is, a NH thermostat with more than one thermostat variable~\cite{martyna1992nose}. This thermostat is global, since the instantaneous value of the temperature is based on a global definition, and non-Galilean invariant. 

The Langevin thermostat~\cite{schneider1978molecular}, which will be discussed in depth in what follows, guarantees ergodicity in all possible cases. Dissipative and noise forces are added to the Hamiltonian to include the effective behaviour of the solvent~\cite{allen2017computer}. Its main disadvantage is that it does not reproduce hydrodynamics~\cite{dunweg1993molecular} because it is not Galilean invariant and does not locally conserve momentum. In order to overcome such a limitation dissipative particle dynamics (DPD) was introduced by Hoogerbrugge and Koelman~\cite{hoogerbrugge1992simulating} and later modified by Espanol and Warren \cite{espanol1995statistical} to satisfy the fluctuation-dissipation theorem. This thermostat represents a modification of the Langevin thermostat. In the DPD thermostat the friction and noise terms are pairwise and act over all pairs of neighbouring particles, \textit{i.e.}, at the local level. In addition, the Galilean invariance is ensured by the fact that the drag force acts on the relative velocity. However, it has several disadvantages. The most important are: {\it (i)} in order to employ large time steps to effectively speed up the simulation, care has to be taken when choosing the right integration algorithm~\cite{lowe1999alternative,besold2000towards,shardlow2003splitting,peters2004elimination,leimkuhler2015numerical}; {\it (ii)} the Schmidt number $Sc=\eta/\rho D$, defined as the ratio between the viscosity $\eta$ and the diffusion constant $D$, is found to be close to 1, while for most common liquids $Sc$ is of the order of $10^{3}$. {\it (iii)} the value of the size of DPD particles, set by $r_c$, cannot be determined \textit{a priori}~\cite{ellero2018everything}. In general, the value of $Sc$ can be improved upon by incorporating a third parameter ($s$) which modifies the weighting function for the dissipative force~\cite{fan2006simulating,phan2017understanding}. As we will discuss below, $Sc$ can be written as a function of $s$, providing us with an efficient way (in computational terms) to increase this number.

In addition, different improvements of the DPD scheme have been developed. For instance, in order to conserve angular momentum, it was proposed to introduce an additional variable~\cite{espanol1997fluid,espanol1998fluid}. Moreover, if the objective is to study chemical reactions where there is a temperature gradient, it is necessary to introduce an additional energy term stemming from the interaction between pairs of particles~\cite{avalos1997dissipative}. Finally, DPD can be also combined with Smoothed Particle Hydrodynamics (SPH)~\cite{liu2010smoothed,wang2016overview} to include the Navier-Stokes equations in the modelling of the solvent particles with the so-called Smoothed Dissipative Particle Dynamics (SDPD)~\cite{espanol2003smoothed,vazquez2016rheology}. It should be noted that DPD can also be used to model an explicit solvent, for example by considering solute and solvent particles with different masses and sizes~\cite{whittle2010dynamic,boromand2017structural}.

More recently, different thermostats implemented as combinations of the above have been proposed. For instance, Stoyanov and Groot~\cite{stoyanov2005molecular} introduced a combination of the Lowe-Andersen (LA) and NH thermostats. A fraction of particles is thermalized with the NH thermostat, while the others are thermalized with the LA. The resulting thermostat is local, Galilean invariant and stable even for large time steps $\Delta t$. However, it presents two main problems: (i) for $\Delta t\rightarrow 0$ it does not converge to the standard DPD, and (ii) it has been shown that it does not sample the canonical ensemble. 

Finally, we mention one last thermostat: the Bussi-Donadio-Parrinello (BDP)~\cite{bussi2007canonical} thermostat. The BDP is a reformulation of the Berendsen thermostat in which the momenta rescaling factor $\alpha$ is computed according to a stochastic evolution of the kinetic energy rather than to a fixed value. Such a change makes the BDP correctly sample the canonical ensemble. 
In its global, rather than local, implementation, hydrodynamics is not correctly reproduced and the specific value of the rescaling frequency $\tau$ does not affect the dynamics~\cite{bussi2008stochastic}.

In this work we focus on non-equilibrium molecular dynamics (NEMD) simulations, aimed in particular to describe rheology experiments.
As in equilibrium, a good control of the temperature is necessary. However, it is also imperative that NEMD simulations faithfully reproduce a realistic dynamics, without any artefacts introduced by the thermostat. This, in turn, requires a detailed assessment of the effects that the thermostat could artificially induce on the behaviour of the system.  To answer this question, we consider a representative model system, i.e. a Lennard-Jones fluid,  under steady shear flow and monitor its behavior under the action of three different thermostats (Langevin, DPD and BDP) for a wide choice of parameters. In recent years both the Langevin~\cite{lander2013crystallization,mohorivc2016two,kohl2017shear} and DPD approaches~\cite{soddemann2003dissipative,zausch2008equilibrium,shrivastav2016yielding} have been used to perform NEMD simulations. The DPD approach has sometimes been the preferred choice because it better reproduces hydrodynamic effects, while the BDP thermostat allows to consider a wider range of shear rates~\cite{zausch2010combined}. However, we will show that a poor choice of the thermostat parameters can negatively affect the dynamic response of  the system under shear, providing a physical picture very far from reality. While there exist direct comparisons between Langevin and DPD thermostats, for instance in the case of coarse-grained bead-spring models for polymers~\cite{pastorino2007comparison}, their absolute or relative performances as a function of the different parameters are seldom discussed in depth. Here we aim to fill this gap, providing a reference test case which can be used as a guidance for choosing a thermostat and its parameters to carry out reliable NEMD simulations. Finally, we note that we are interested in bulk systems, and hence we will consider systems where periodic boundary conditions are enforced, as we will describe below. However, NEMD simulations can also be performed with wall boundaries that constrain the system in a well-defined geometry. The walls can then be moved to reproduce an oscillatory or steady shear to drive the system away from equilibrium~\cite{doyle1997rheology,yong2010investigating}. In addition, it is also possible to play with the nature of the wall in order to study slip effects~\cite{barrat1999large,priezjev2006influence,niavarani2010modeling}. Also in this case, different thermostating strategies are available~\cite{yong2013thermostats}.

We note on passing that the integration algorithms are also an important point to consider~\cite{leimkuhler2015numerical}, even if we do not explicitly focus on this aspect.

The paper is organised as follows. In Sec.~\ref{sec:methods} we introduce the simulation details and describe in detail the three thermostats studied in this work. We then discuss the different observables which we have measured to study the behaviour of the thermostats. In Sec.~\ref{sec:results} the effects of the different thermostats are shown under both equilibrium and shear flow conditions. In Sec.~\ref{sec:conclusions} we draw our conclusions.

\section{Simulations details}
\label{sec:methods}

\subsection{Equilibrium}
We study a system composed of $N=2000$ monodisperse particles of mass $m$ and size $\sigma$ interacting through a Lennard-Jones potential
\begin{equation}
	\label{eq:LJ}
	V\left(r\right)=4\epsilon\left[\left(\frac{\sigma}{r}\right)^{12}-\left(\frac{\sigma}{r}\right)^{6}\right]
\end{equation}

\noindent
where $\epsilon$ controls the depth of the potential. The parameters $\sigma$ and $\epsilon$ are chosen as units of length and energy, respectively. We also set $k_{B}=1$. The potential is cut at $r_{c}=2.5\sigma$. We fix the number density of the system to $\rho = 0.844$ and  study two different temperatures, $T = 1.5$ and $T=0.722$,  corresponding to liquid-like states in the supercritical region and close to the triple point, respectively. 
In order to reduce the numerical errors intrinsic to integration schemes~\cite{leimkuhler2015numerical,leimkuhler2016pairwise,shang2017assessing}, we fix the time step to $\Delta t=0.002$ for both equilibrium and NEMD simulations.

\subsection{Steady shear}

Shear flow is applied by using Lees-Edwards boundary conditions \cite{lees1972computer}, where  different layers of image boxes in the direction of the flow gradient are considered as moving with a shear velocity $v_{s}$. Hence, the shear rate is $\dot\gamma=v_{s}/L$, where $L$ 
is the length of the simulation box. The modified periodic boundary conditions thus impose a time-dependent linear shear flow in the $x$ direction such that the shear gradient is parallel to $z$ and the vorticity is along the $y$ direction. The flow velocity $\mathbf{u} = \dot\gamma z\mathbf{x}$ depends linearly on $z$ (planar Couette flow) and is zero in the centre of the channel ($z = L/2$). In this work we use $\dot\gamma=0.01$ and $\dot\gamma=0.1$. We notice that these conditions can be also investigated by solving the SLLOD equations of motion~\cite{morriss2013statistical}, for which several thermostats have been adapted~\cite{zhang1999kinetic,pan2005operator}. However, this algorithm has not been used in this work.

In general, the energy injected by the shear flow needs to be dissipated by the thermostat. The mechanism through which this happens affects the flow profile. There are two classes of thermostats that allow to dissipate the extra energy: {\it profile-unbiased thermostats} (PUT), which allow the velocity profile to emerge as a characteristic response of the system and {\it profile-biased thermostats} (PBT), which enforce a fixed streaming velocity profile~\cite{todd2007homogeneous}. In the following we will present and use thermostats of both kinds.

\subsection{Langevin thermostat}

The equations of motion for the Langevin dynamics are:

\begin{equation}
\begin{split}
	\label{eq:Langevin}
	& m_{i}\mathbf{\dot{r}}_{i}=\mathbf{v}_{i} \\
	& m_{i}\mathbf{\ddot{r}}_{i}=\sum_{j\left(\neq i\right)}\mathbf{F}_{ij}^{C}+\mathbf{F}_{i}^{R}+\mathbf{F}_{i}^{D}
\end{split}
\end{equation}

\noindent
where the first term is the usual conservative pairwise force, $\mathbf{F}_{i}^{D}=-\xi m\mathbf{v}_{i}$ is the dissipative (drag) force with friction constant $\xi$ and $\mathbf{F}_{i}^{R}$ is the random force due to the thermal motion of the bath particles, modelled as a stochastic white noise with zero mean and variance 

\begin{equation}
	\label{eq:Lang_R}
	\left\langle\mathbf{F}_{i}^{R}\left(t\right)\cdot\mathbf{F}_{j}^{R}\left(t'\right) \right\rangle = \sqrt{2k_{B}T\xi}\delta_{ij}\delta\left(t-t'\right).
\end{equation}

\noindent
The variance of the stochastic force is set by the fluctuation-dissipation theorem. The combination of dissipative and random forces is used to represent the effect of the solvent on the system.  The parameter $\xi$ effectively controls the viscosity of such fictitious solvent. Depending on its value, the system can be found either in the Brownian (overdamped Langevin dynamics) or in the Langevin regime (underdamped Langevin dynamics). 
The integration of Eq.\eqref{eq:Langevin} is carried out with the self-adaptive OVRVO~\cite{sivak2014time} scheme: if $\xi\Delta t \gg 1$, the algorithm reduces to the Euler-Maruyama method~\cite{ermak1978brownian}, whereas for $\xi\Delta t \ll 1$ the algorithm is equivalent to the Langevin method~\cite{brunger1984stochastic}. This algorithm is a profile-unbiased thermostat, which is suitable for both equilibrium and non-equilibrium dynamics.

We further implement two variants of the Langevin thermostat. The first one consists in turning off the conservative and dissipative forces along one or two directions. This modification has been sometimes shown to be able to correct some spurious artefacts appearing under non-equilibrium dynamics~ \cite{soddemann2003dissipative}. In what follows the Langevin thermostat acting on one, two and three directions will be indicated with $\xi_{\rm y}$, $\xi_{\rm yz}$ and $\xi_{\rm xyz}$, respectively. In the second variant, the drag force acts on the peculiar velocity, that is, on the difference between the absolute velocity $\mathbf{v}_{i}$, and the streaming velocity field or shear flow $\mathbf{u}$, rather than on the absolute velocity. With this modification Eq.~\eqref{eq:Langevin} becomes

\begin{equation}
	\label{eq:Peculiar}
	m_{i}\mathbf{\ddot{r}}_{i}=\sum_{j\left(\neq i\right)}\mathbf{F}_{ij}^{C}-\xi\left(\mathbf{p}_{i}-m\mathbf{u}_{i}	\right)+\mathbf{F}_{i}^{R}	
\end{equation}

\noindent
This enforces the velocity profile to be linear (profile-biased conditions). We integrate Eq.~\eqref{eq:Peculiar} with the so-called BAOAB \cite{shang2017assessing} algorithm. The use of OVRVO and BAOAB algorithms do not present any difference in equilibrium, when $\mathbf{u}=0$. In the following we will use the symbol $\xi_{\rm pec}$ to refer to this version of the Langevin thermostat, acting on the `peculiar' velocity.

\subsection{Dissipative Particle Dynamics}

The DPD thermostat locally conserves momentum because all forces act between pairs of particles. It is also Galilean invariant because the dissipative force acts on relative velocities only. Since it does not explicitly enforce a linear velocity profile, the DPD thermostat is a PUT. The equations of motion are written as:
\begin{equation}
\begin{split}
	\label{eq:DPD}
	& m_{i}\mathbf{\dot{r}}_{i}=\mathbf{v}_{i} \\
	& m_{i}\mathbf{\ddot{r}}_{i}=\sum_{j \neq i}\left[\mathbf{F}_{ij}^{C}+\mathbf{F}_{ij}^{R}+\mathbf{F}_{ij}^{D}\right]
\end{split}
\end{equation}

\noindent
where we use the same notation as above and the pairwise random and dissipative forces are given by

\begin{equation}
	\label{eq:DPD_R}
	\mathbf{F}_{ij}^{R}=\sqrt{2k_{B}T\xi} w^{R}\left(r_{ij}\right)\theta_{ij}\hat{\mathbf{r}}_{ij}
\end{equation}
\begin{equation}
	\label{eq:DPD_D}
	\mathbf{F}_{ij}^{D}=-\xi w^{D}\left(r_{ij}\right)\left[\hat{\mathbf{r}}_{ij}\cdot\mathbf{v}_{ij}\right]\hat{\mathbf{r}}_{ij}
\end{equation}

\noindent
where $\mathbf{v}_{ij}=\mathbf{v}_{i}-\mathbf{v}_{j}$ and $\mathbf{r}_{ij}=\mathbf{r}_{i}-\mathbf{r}_{j}$ are the relative velocities and distances of particles $i$ and $j$, respectively, and $\hat{\mathbf{r}}_{ij}=\mathbf{r}_{ij}/r_{ij}$. 
The variable $\theta_{ij}=\theta_{ji}$ is a Gaussian noise term with zero mean and variance given by
\begin{equation}
	\label{eq:DPD_Var}
	\left\langle \theta_{ij}\left(t\right)\theta_{kl}\left(t'\right) \right\rangle = \left(\delta_{ik}\delta_{jl}+\delta_{il}\delta_{jk}\right)\delta\left(t-t'\right).
\end{equation}

The functions $w^{D}\left(\mathbf{r}_{ij}\right)$ and $w^{R}\left(\mathbf{r}_{ij}\right)$ are weight functions. They are related by the stationary solution of the Fokker-Planck equation where the dissipation-fluctuation theorem is satisfied through the relation~\cite{espanol1995statistical}

\begin{equation}
	\label{eq:Diss-Fluc}
	w^{D}\left(r_{ij}\right)=\left[w^{R}\left(r_{ij}\right)\right]^{2}
\end{equation}

\noindent
Consequently, one of these two weight functions can be chosen freely. Fan \textit{et al.} introduced a generalized weighting function for the dissipative force~\cite{fan2006simulating}
\begin{equation}
  \label{eq:weight}
  \omega^{D}\left(r_{ij}\right)=\left[\omega^{R}\left(r_{ij}\right)\right]^{2}=\left\{ \begin{array}{c}
  \left(1-\frac{r_{ij}}{r_{c}}\right)^{s}\qquad r_{ij}<r_{c}\\
  0\qquad\qquad\quad\quad r_{ij}\geq r_{c}.
  \end{array}\right.
\end{equation}
\noindent where $s$ and $r_c$ are the exponent and the cutoff radius of the weighting function, respectively. The latter parameter defines the size of DPD particles~\cite{ellero2018everything}. Taking $r_{c}=1$ and $s=2$ we recover the conventional DPD algorithm, as obtained by Groot and Warren~\cite{groot1997dissipative}. 

The three parameters of the thermostat, i.e. $s$, $r_c$ and $\xi$ determine the transport properties of the system~\cite{fan2006simulating,moshfegh2015dissipative}. Indeed, in Ref.~\cite{phan2017understanding} it is shown that both the shear viscosity $\eta$ and the constant diffusion $D$, and thus also the Schmidt number, all depend on the three DPD parameters. However, the value of $r_c$ greatly affects the computational cost, since it controls the number of pairs of particles that enters into the thermostating procedure. Therefore, changes to $s$ or $\xi$ present a more efficient way to obtain a more realistic value of $Sc$. Furthermore, as it will be shown below, they play an important role not only in the control of the temperature, but also in the generic dynamic response of the system under shear.

The DPD equations of motion can be integrated by different schemes~\cite{leimkuhler2015numerical}, the two most famous ones being the Lowe-Andersen \cite{lowe1999alternative} and Peters~\cite{peters2004elimination} algorithms. The first one is inspired by the Andersen thermostat, but instead of thermalizing the velocity of individual particles, it thermalizes the relative velocities of two neighbouring particles with a probability $P=\Gamma \Delta t$. However, for $\Delta t\rightarrow0$ it does not converge to the standard DPD system. This issue is resolved by adopting the Peters scheme, as done in this work.
Two observations are in order. First of all, we have tested that the thermalization does not need to follow a random order, as suggested in Ref.~\cite{peters2004elimination}, but can be carried out sequentially, as also mentioned in Ref.~\cite{leimkuhler2015numerical}. Secondly, it has been suggested that, for the sake of efficiency, Gaussian random numbers of zero mean and unitary variance can be well-approximated by using a uniform distribution between $\pm\sqrt{3}$ \cite{dunweg1991brownian}. As shown below (see Fig~\ref{fig:VAC_DPD}(b)), this is not always the case, as there exist combinations of parameters for which non-Gaussian random numbers yield the wrong temperature.

\subsection{Bussi-Donadio-Parrinello (BDP) thermostat}

The BDP thermostat is similar in spirit to the Berendsen thermostat, because the velocities of the particles are rescaled by a factor $\alpha$~ \cite{berendsen1984molecular}. In both cases $\alpha = \sqrt{\frac{K_t}{K}}$, where $K_t$ is the target value of the kinetic energy at time $t$ and $K$ is the instantaneous kinetic energy. However, in the BDP case $K_{t}$ is not constant but evolves stochastically according to the canonical distribution of the kinetic energy~\cite{bussi2007canonical}. Ref.~\cite{bussi2007canonical} gives an explicit expression for the rescaling factor, which reads
\begin{multline}
	\label{eq:alpha}
	\alpha^2=e^{-\Delta t/\tau}+\frac{\bar K}{N_{f}K}\left(1-e^{-\Delta t/\tau}\right)\left(R_{1}^{2}+\sum_{i+2}^{N_f}R_{i}^{2}\right)+ \\
	2R_{1}e^{-\Delta t/\tau}\sqrt{\frac{\bar K}{N_{f}K}\left(1-e^{-\Delta t/\tau}\right)}
\end{multline}

\noindent
where ${\bar K}=N_{f}/2k_{B}T$ is the target temperature, $N_f$ is the number of degrees of freedom, $R_i$'s are independent random numbers extracted from a Gaussian distribution with zero mean and unitary variance and the parameter $\tau$, which is defined as $\tau=\left(2\xi\right)^{-1}$ (dimensions of time), determines the time scale of the thermostat. Note that, in the non-local implementation of the BDP that we use here, the parameter $\tau$ does not truly represent the effect of the solvent\cite{bussi2008stochastic}, and thus a change in $\tau$ has basically no effect on the dynamics of the system.

The equations of motion are integrated with the velocity Verlet method~\cite{allen2017computer}. The BDP thermostat is a PBT since it acts on the peculiar velocities of the particles, which are rescaled by the factor $\alpha$ after a specified number of time steps.

\subsection{Observables}
\label{subsec:observables}

We run simulations under equilibrium and non-equilibrium conditions. We investigate the dynamics of the system by looking at the mean squared displacement $\left\langle  r^{2}\left(t\right)\right\rangle$ from which we extract the diffusion coefficient $D$, defined as
\begin{equation}
	\label{eq:Diffusion}
	D = \lim_{t\to\infty} \frac{\left\langle  r^{2}\left(t\right)\right\rangle}{6t}.
\end{equation}

\noindent
In rheology experiments and simulations, this parameter is fundamental to describe the competition between diffusion and shear flow. In equilibrium, the dynamics of a fluid is governed by Brownian (diffusive) dynamics. In the presence of a shear flow an anisotropic microstructure, originated from the competition between diffusion and shear effects, appears in the stationary state. While Brownian dynamics tends to restore the equilibrium of the system, the shear flow tends to distort its structure. The competition between the two regimes is encoded in the Peclet number $Pe=\tau_B/\tau_S = \sigma^2\dot\gamma / (4D)$, where $\tau_B$ and $\tau_S$ are the characteristic times of diffusion and shear flow, respectively~\cite{cloitre2010high}. 

We also compute the zero-shear viscosity $\eta$ by using the Green-Kubo relation~\cite{hansen1990theory},

\begin{equation}
	\label{Viscosity}
	\eta\equiv\int_{0}^\infty C_{\sigma\sigma}(t) dt=\frac{\beta}{3V}\int_{0}^\infty \sum_{\alpha<\beta} \left\langle \sigma^{\alpha\beta}\left(t\right)\sigma^{\alpha\beta}\left(0\right) \right\rangle dt
\end{equation}

\noindent
where $\beta=1/k_{B}T$, $V$ is the volume of the simulation box, $\langle \cdots \rangle$ indicates an average over initial conditions and the microscopic stress tensor $\sigma^{\alpha\beta}$ is defined as

\begin{equation}
	\label{Stress Tensor}
	\sigma^{\alpha\beta}=\sum_{i=1}^{N}m_{i}v_{i\alpha}v_{i\beta}+\sum_{i<j}^{N}\frac{r_{ij\alpha}r_{ij\beta}}{r_{ij}}F\left(r_{ij}\right)
\end{equation}

\noindent
where $v_{i\alpha}$ is the $\alpha$-th component of the velocity of particle $i$, $F\left(r_{ij}\right)$ is the force between particles $i$ and $j$. Since numerical estimates of the stress autocorrelation function $C_{\sigma\sigma}(t)$ are, especially at high friction, very noisy, we follow previous work and fit $C_{\sigma\sigma}(t)$ with the phenomenological expression provided in Ref.~\cite{puertas2007viscoelasticity}.

Finally, we analyse how particles exchange momentum with their surroundings through the velocity autocorrelation function, defined as 
\begin{equation}
Z(t)=\frac{1}{3}\left\langle \bold{v}\left(t\right)\cdot \bold{v}\left(0\right)\right\rangle.
\end{equation}
The decay of this function provides information on how different thermostats make the system's dynamics decorrelate over time.

When subject to shear flow, the local properties of the system, both static and dynamic, change along the gradient direction $z$. It is therefore useful to divide the simulation volume into layers of thickness $\sigma$ and area $L\times L$ perpendicular to the gradient direction. We thus consider: (i) the local temperature $T(z)$, extracted from the kinetic energy computed from the components of the velocity which are perpendicular to the shear direction,  (ii) the local density $\rho(z)$, defined as the number of particles in the layer divided by the volume of the layer ($\sigma L^2$) and (iii) the profile of the $x$-component of the velocity $v_x(z)$, defined as the average of the velocity along $x$ over all particles in the layer. Regarding the latter observable, in equilibrium there is no preferred direction and hence $\left\langle v_{x}\left(z\right) \right\rangle\sim0$ in all layers. By contrast, the velocity profile of a liquid under shear flow is linear with $x$ and $\left\langle v_{x}\left(\pm L/2\right) \right\rangle=\pm\dot\gamma L/2$. If this is not the case then the velocity profile will exhibit anomalies, \textit{i.e.}, the appearance of regions where the deformation of the system is not uniform with respect to the external field. As we will show below, this is related to inhomogeneities in the density profile, which might indicate a non-physical behaviour.

\begin{figure*}[t!]
\centering \includegraphics[width=1.0\textwidth]{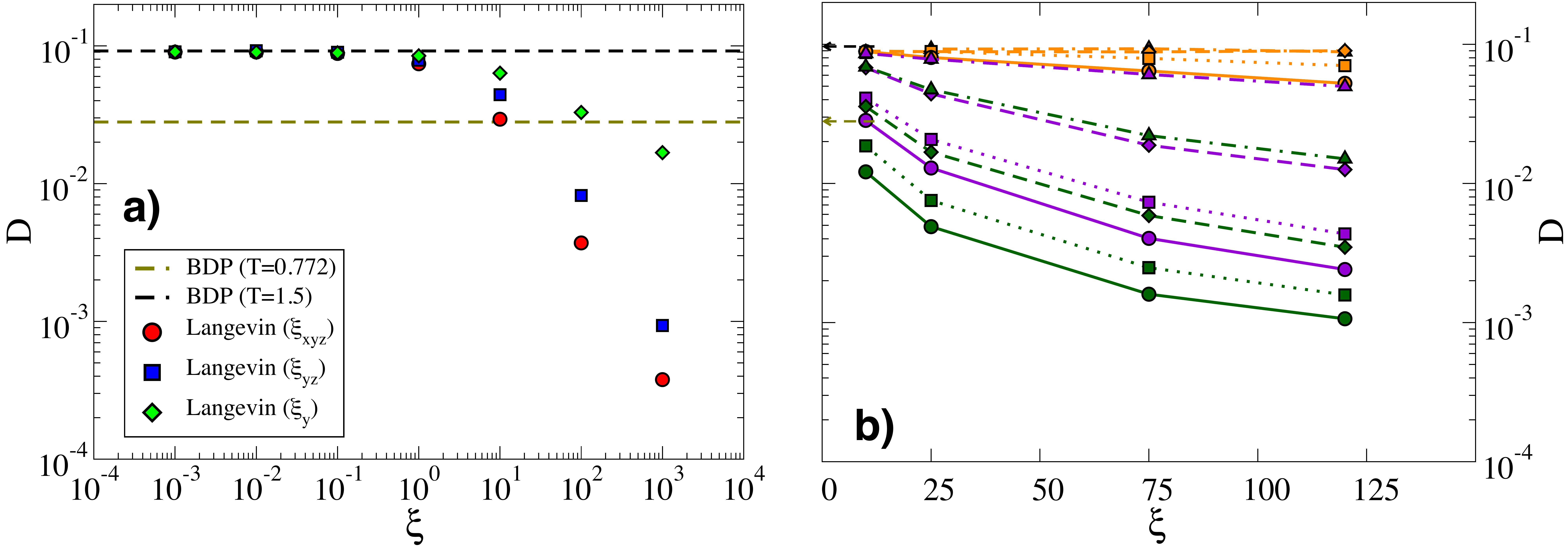}
\caption{Diffusion coefficient $D$ for the system in equilibrium at $T=1.5$ and $\rho=0.844$. a) $D$ calculated with the Langevin thermostat acting along one, two and three directions. For comparison $D$ calculated with the BDP thermostat is also shown at both $T = 1.5$ and $T = 0.772$ (see Ref. \cite{bussi2008stochastic}). b) $D$ calculated with DPD thermostat for different sets of parameter. Different symbols correspond to different values of the exponent: $s=0.25$ (circles), $s=0.50$ (squares), $s=1.0$ (diamonds) and $s=2.0$ (triangles), while different colours indicate different cut-off values: $r_{c}=1.12$ (orange), $r_{c}=1.54$ (purple) and $r_{c}=1.88$ (dark green).}
\label{fig:MSD}
\end{figure*}

\begin{figure*}[t!]
\centering \includegraphics[width=1.0\textwidth]{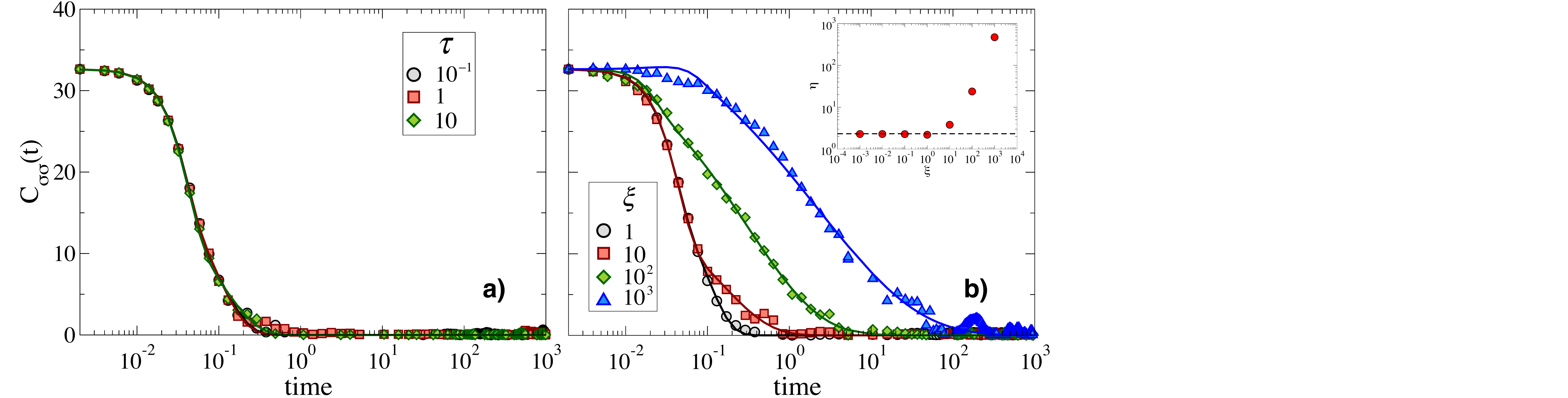}
\caption{Stress correlation function $C_{\sigma\sigma}\left(t\right)$ for a) the BDP thermostat and b) the Langevin thermostat $\xi_{xyz}$. Symbols represent the numerical $C_{\sigma\sigma}\left(t\right)$, whereas solid lines are fits~\cite{puertas2007viscoelasticity}. {\it Inset:} The dashed line shows the BDP reference viscosity, whereas symbols indicate the $\xi_{xyz}$ Langevin viscosity.}
\label{fig:Viscosity1}
\end{figure*}

\begin{figure*}[t!]
\centering \includegraphics[width=1.0\textwidth]{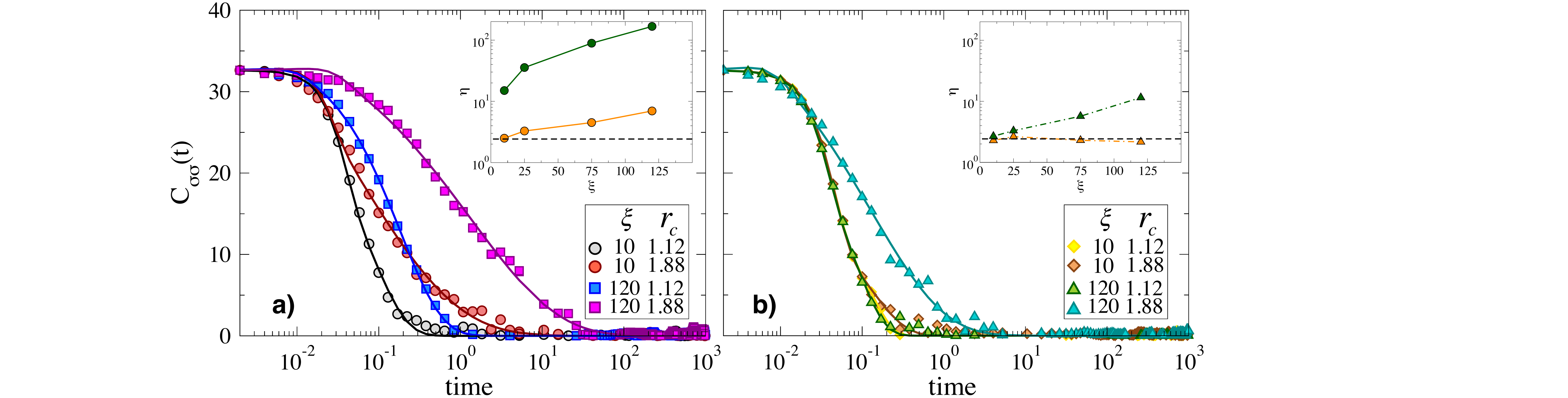}
\caption{Stress correlation function $C_{\sigma\sigma}\left(t\right)$ for the DPD thermostat for a) $s=0.25$ and b) $s=2.0$. Symbols represent the numerical $C_{\sigma\sigma}\left(t\right)$, whereas solid lines are fits~\cite{puertas2007viscoelasticity}. {\it Insets:} The dashed line shows the BDP reference viscosity, whereas symbols indicate the DPD viscosity.}
\label{fig:Viscosity2}
\end{figure*}

\begin{figure*}[t!]
\centering \includegraphics[width=0.9\textwidth]{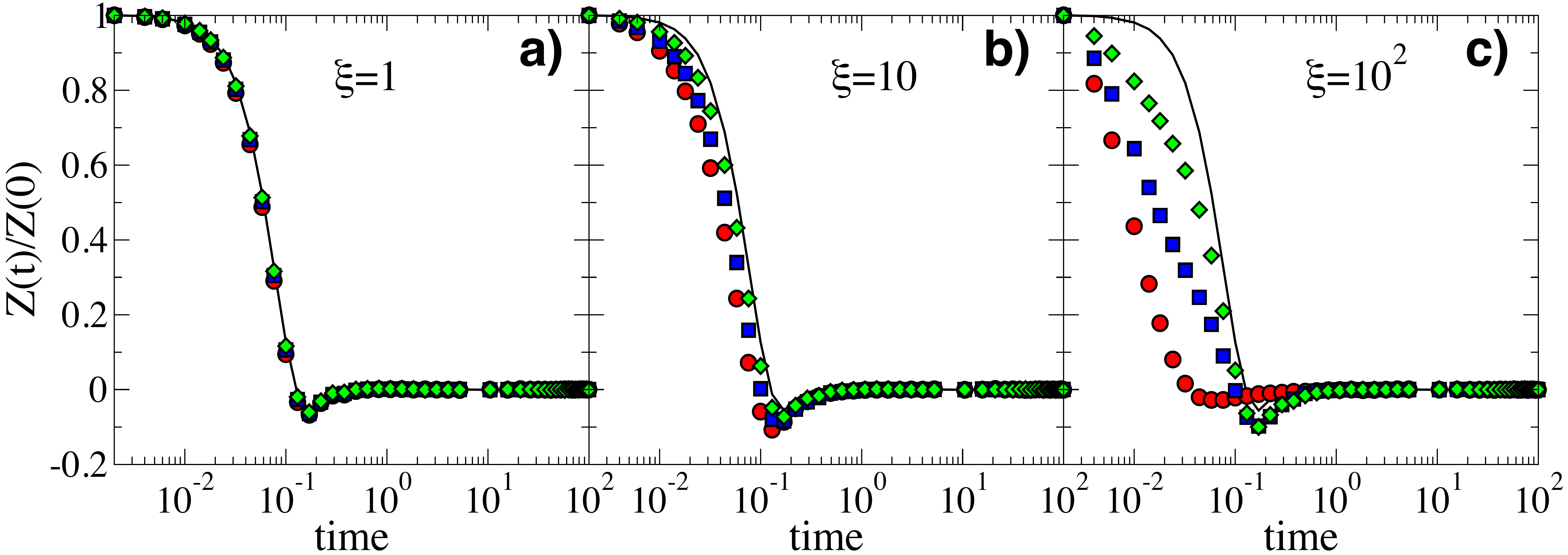}
\caption{Velocity-autocorrelation function $Z\left(t\right)$ (normalized by its zero-time value) for the system in equilibrium at $T=1.5$ and $\rho=0.844$ with the Langevin thermostat acting in three (red circles), two (blue squares) and one directions (green diamonds) and three values of the friction constant: a) $\xi=1$, b) $\xi=10$ and c) $\xi=10^{2}$. The black (reference) curve is obtained with the BDP thermostat.}
\label{fig:VAC_LAN}
\end{figure*}

\begin{figure*}[t!]
\centering \includegraphics[width=0.9\textwidth]{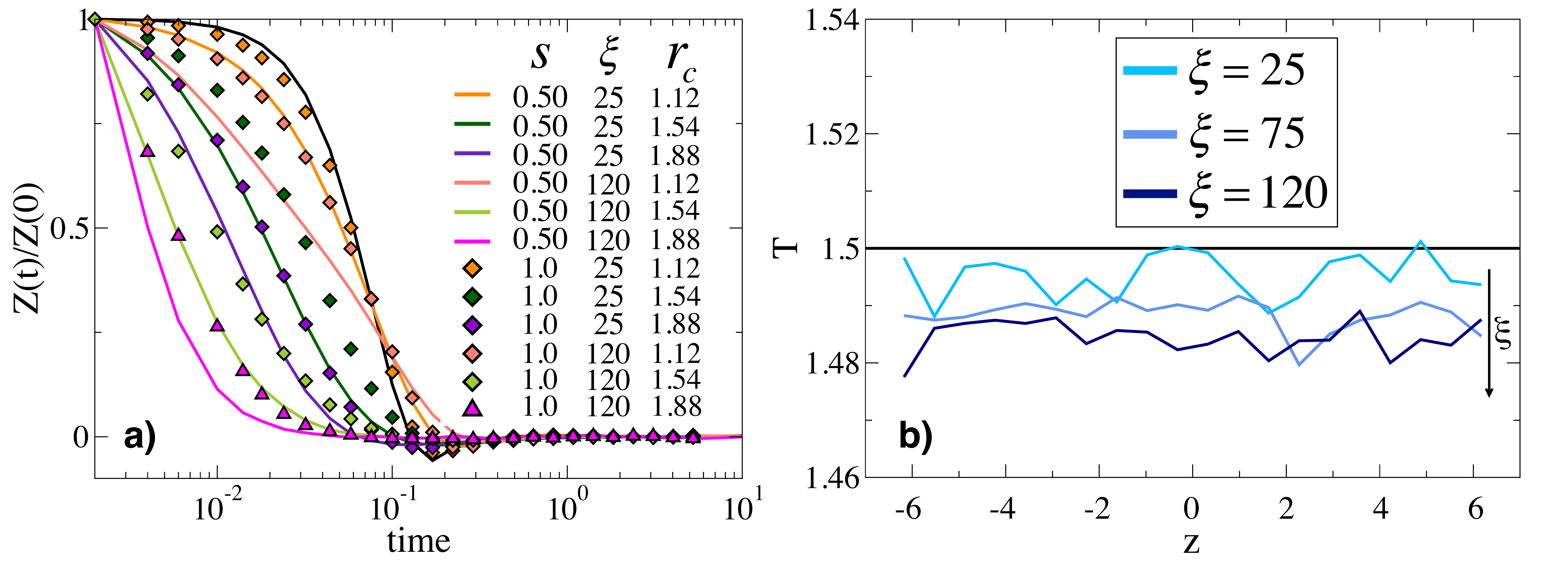}
\caption{(a) Velocity-autocorrelation function $Z\left(t\right)$  (normalized by its zero-time value) for the system in equilibrium at $T=1.5$ and $\rho=0.844$ with the DPD thermostat for several combinations of the parameters. The black (reference) curve is obtained with the BDP thermostat. (b) Temperature profile obtained with the DPD thermostat using $s=0.25$, $r_c=1.12$ and various values of $\xi$ extracting random numbers from a uniform distribution~\cite{dunweg1991brownian}. Clearly, the temperature decreases as $\xi$ increases.}
\label{fig:VAC_DPD}
\end{figure*}

\section{Results}
\label{sec:results}

\subsection{Equilibrium}

In this section we focus on the system at $T = 1.5$ in equilibrium. We start by calculating the diffusion coefficient $D$,  reported in Fig~\ref{fig:MSD}(a). First of all, we note that $D$ obtained via BDP simulations is independent of $\tau$ (and hence of $\xi = (2\tau)^{-1}$)~\cite{bussi2008stochastic}. Since the same holds true for all other observables investigated, in what follows we use the BDP data as reference. We have checked that the same reference curves are obtained if, at zero shear, no thermostats are employed, that is, if the simulations are performed in the {\it NVE} ensemble. For both Langevin and DPD, $D$ is a monotonically decreasing function of $\xi$. For the Langevin thermostat, the fewer directions the friction acts on, the larger is the diffusion coefficient. However, when $\xi\leq10^{-2}$ Brownian effects are small and $D$ always tends to a plateau. For the DPD thermostat, the effect of the two other parameters is also monotonic: $D$ increases with $s$ and decreases with $r_c$. The dependence on $s$ can be explained by considering that lower values of $s$ cause higher correlations between the drag and random forces in the equations of motion, thereby making the system decorrelate faster. By contrast, the increase of $D$ with $r_c$ is related to the fact that the larger the cut-off, the more particles are included in the thermalization procedure. Similar effects have been observed with DPD-related thermostats~\cite{fan2006simulating,junghans2008transport,qian2009effective}. We note that the range of values of $D$ is comparable for both thermostats, indicating that they share the same nature. A variation of two orders of magnitude in $D$ is observed in the explored range of thermostat parameters.

The choice of the parameters also affects the viscosity which, for the systems investigated here, gives similar information as $D$. For instance, the BDP thermostat shows that $C_{\sigma\sigma}\left(t\right)$ is independent of $\tau$, which translates to a constant viscosity, as shown in Fig.~\ref{fig:Viscosity1}(a). With respect to the Langevin thermostat, in the same range where $D$ is constant, $\eta$ is also constant (see Fig.~\ref{fig:Viscosity1}(b)). Finally, Fig.~\ref{fig:Viscosity2}(a) and (b) shows that, for the DPD thermostat, the value of $\eta$ displays a strong dependence on $s$. In particular, smaller values of $s$ induce larger correlations on the microscopic internal stress, effectively generating larger viscosities.

Figure~\ref{fig:VAC_LAN} shows $Z\left(t\right)$ (normalized by its zero-time value) for the Langevin thermostat acting on three, two and one directions for three different values of $\xi$. At low friction ($\xi = 1$), the three curves fall on top of the reference (BDP) one. However, upon increasing $\xi$ we see that the correlations are removed earlier for the Langevin thermostat that acts on all three spatial directions. By removing a direction of thermalization, the overall friction coefficient is effectively reduced. This is in agreement with the behaviour observed for $D$ (see Fig~\ref{fig:MSD}(a)).

The velocity autocorrelation functions for the DPD thermostat are shown in Fig~\ref{fig:VAC_DPD}(a). The dependence of its decay on $s$ and $r_c$ is directly linked to the effect that these two parameters have on $D$ (see Fig.\ref{fig:MSD}(b)), and the effect of $\xi$ is the same as in the Langevin thermostat. Overall, the combined effect of the three DPD thermostat parameters is to tune the viscosity of the system in a finer way with respect to Langevin dynamics. 

A useful tip concerns the use of random numbers extracted from uniform distributions, which allows to speed up the computation\cite{dunweg1991brownian}. We find that this might, in principle, produce the wrong temperature profile, depending on the parameters used. Indeed, Figure~\ref{fig:VAC_DPD}(b) shows an example that was generated with a DPD thermostat with $s = 0.25$ and $r_c = 1.12$. For this choice of parameters, the temperature clearly decreases with increasing $\xi$, a problem that is avoided by using a Gaussian distribution for the random numbers.

We conclude by noting that, at short times ($t\rightarrow0$), the slope of $Z\left(t\right)$ should tend to zero, as seen for the reference curve. However, when $\xi$ increases the slope of the autocorrelation function at $t = 0$ becomes more and more negative for both DPD and Langevin thermostats. This is due to the local nature of the thermostating scheme, as discussed in Refs.~\cite{koopman2006advantages,bussi2009isothermal}.

\subsection{Steady shear with Langevin dynamics}

\begin{figure*}[t!]
\centering \includegraphics[width=0.9\textwidth]{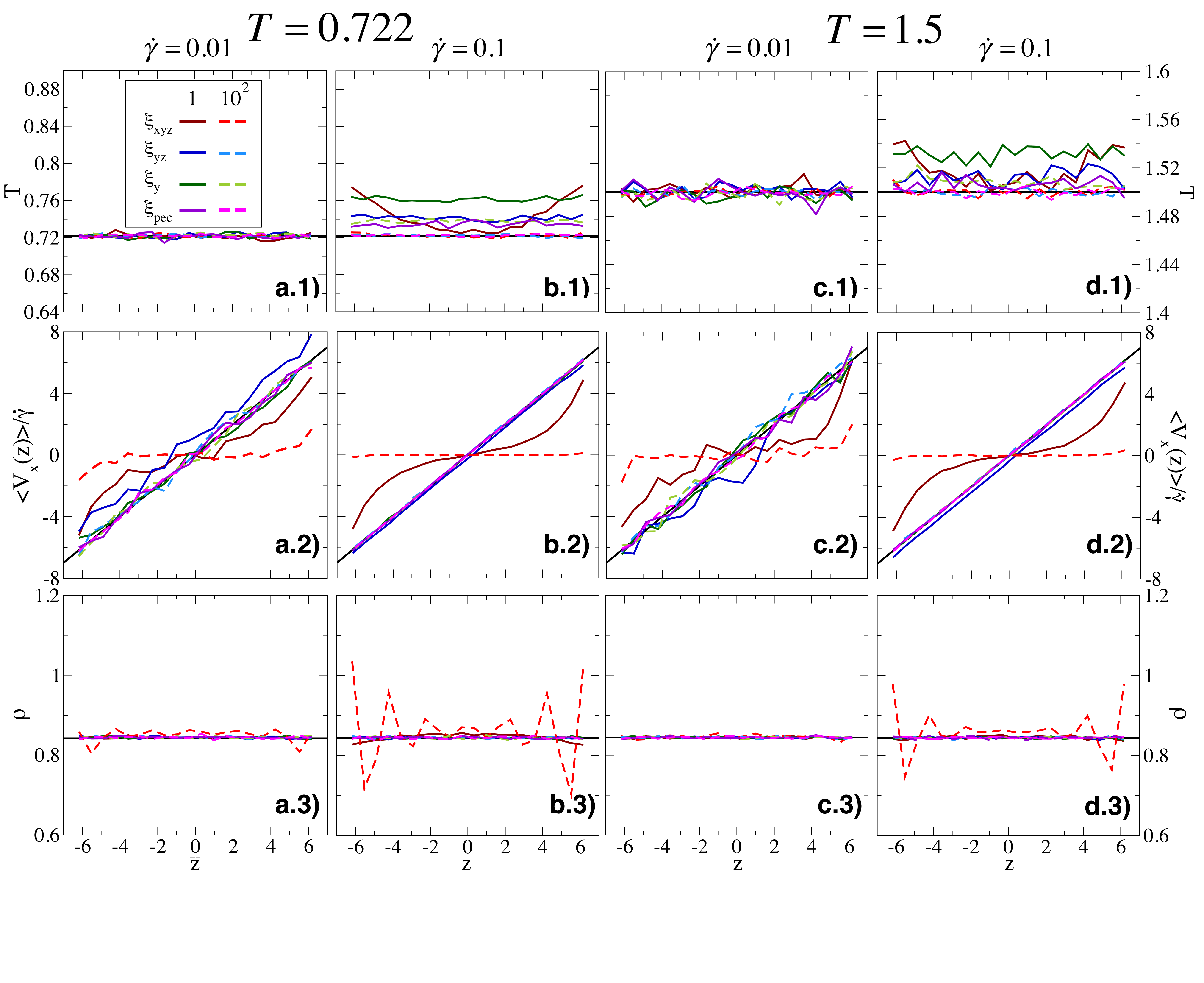}
\caption{Temperature (top), velocity (middle) and density (bottom) profiles computed with the Langevin thermostat under shear flow, for two values of $T$ and $\dot\gamma$. Here $\xi_{xyz},\xi_{yz}, \xi_y$ indicate thermalization acting on three, two, one direction respectively, while $\xi_{pec}$ refers to the thermostat acting on the peculiar velocities. The black lines are the reference curves.}
\label{fig:Prof_Lan}
\end{figure*}

\begin{figure*}[t!]
\centering \includegraphics[width=0.9\textwidth]{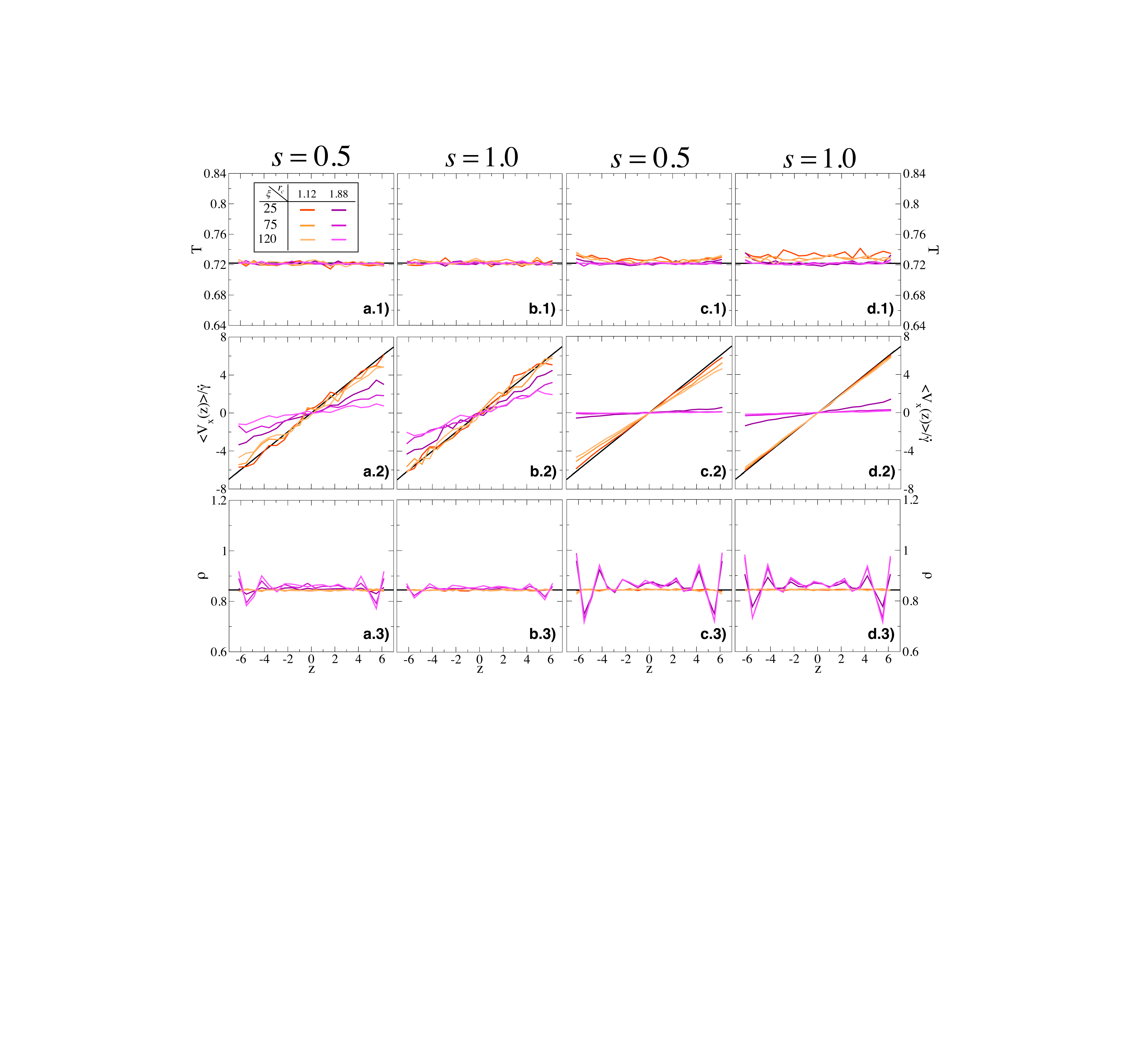}
\caption{Temperature (top), velocity (middle) and density (bottom) profiles computed with the DPD thermostat under shear flow for $T=0.722$, (a-b) $\dot\gamma=0.01$ and (c-d) $\dot\gamma=0.1$.  The black lines are the reference curves.}
\label{fig:Prof_DPD_T0.772_0.01}
\end{figure*}

We now focus on the system under steady shear using a Langevin thermostat. Figure~\ref{fig:Prof_Lan} shows $T(z)$, $v_x(z)$ and $\rho(z)$ for both studied values of $T$ and $\dot\gamma$ and for different values of the friction constant. For all values of $\xi$ the temperature is constant at low shear rate (see Fig~\ref{fig:Prof_Lan}(a.1) and (c.1)), regardless of the number of spatial directions to which thermalization is applied. However, upon increasing $\dot\gamma$ the temperature profile remains independent of $z$ only for the three-directions Langevin thermostat and for large enough values of $\xi$. The density profile (see Fig~\ref{fig:Prof_Lan}(a.3-d.3)) exhibits similar trends. By contrast, the velocity profile $v_x(z)$ has a more complex behaviour, as it is linear only in some cases. From the plots it is clear that a flat $T(z)$ does not imply a linear velocity profile and \textit{vice versa}. Indeed, a constant $T(z)$ can be obtained by increasing $\xi$, but the consequent increase in the fluid viscosity can cause the appearance of a non-linear velocity profile and unrealistic inhomogeneities in the density profile. These inhomogeneities are stronger close to the border of the simulation box, where the flow velocity is higher ($\pm\dot{\gamma}\frac{L}{2}$). The inhomogeneities become more pronounced upon increasing $\dot\gamma$, indicating that the Langevin dynamics cannot cope with the increased stress, or upon decreasing $T$, when more stable particle aggregates tend to form. These results clearly show that a correct control of the temperature does not guarantee that other fundamental observables, such as the velocity and density profiles, are correctly reproduced. 

Figure~\ref{fig:Prof_Lan} also shows that this problem can be only partially mitigated by turning off the thermostat along one (the flow) or two (the flow and gradient) directions: the modified thermostats work at low $\dot\gamma$ only. However, we do not recommend using this technique as it introduces additional spatial inhomogeneities in the system~\cite{morriss2013statistical,leimkuhler2016pairwise}. Better performances can be obtained instead by coupling the system to a Langevin thermostat that acts on the peculiar velocity. In this case, indicated in Figure~\ref{fig:Prof_Lan}  as $\xi_{pec}$,  a linear velocity profile is, by construction, always recovered. Furthermore, the density profile shows a perfect homogeneity for any value of $\xi$. However, in even more extreme cases this modification might still not be enough to ensure a correct thermalisation, and hence a flat temperature profile, as for too large values of the shear rate, the thermostat might not be able to dissipate the extra kinetic energy. However, with the peculiar Langevin thermostat the linear velocity profile is an \textit{enforced} rather than an emergent property, meaning that, for instance, shear banding would not be observed even in cases where it should be present~\cite{soddemann2003dissipative,shrivastav2016yielding,kang2017non}. For the same reason, the peculiar Langevin thermostat could hide short-time dynamic heterogeneities which appear in glasses~\cite{shrivastav2016yielding}.

\begin{figure*}[t!]
\centering \includegraphics[width=0.9\textwidth]{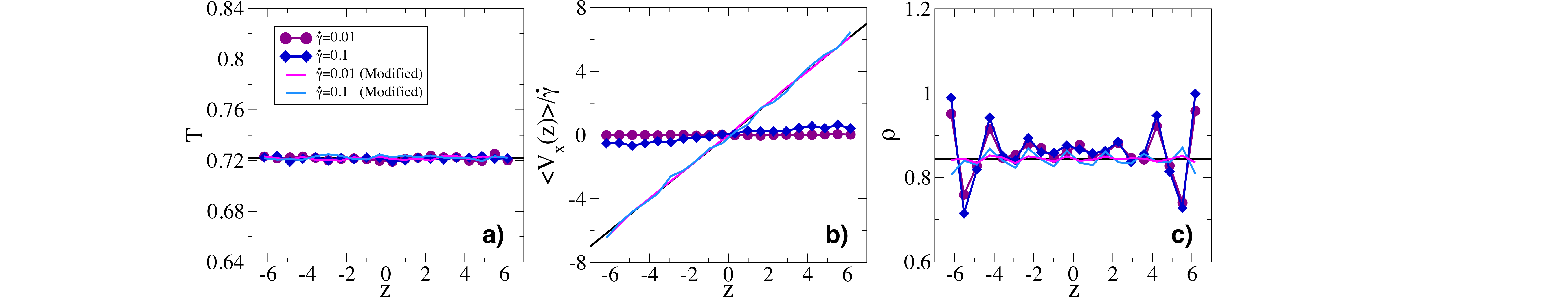}
\caption{(a) Temperature, (b) velocity and (c) density profiles for a DPD thermostat with $s=0.25$, $\xi=120$ and $r_{c}=1.88$ with and without the improved treatment of the periodic boundary conditions (see text for details). The black lines are the reference curves.}
\label{fig:Prof_DPD_MOD}
\end{figure*}

\begin{figure*}[t!]
\centering \includegraphics[width=0.9\textwidth]{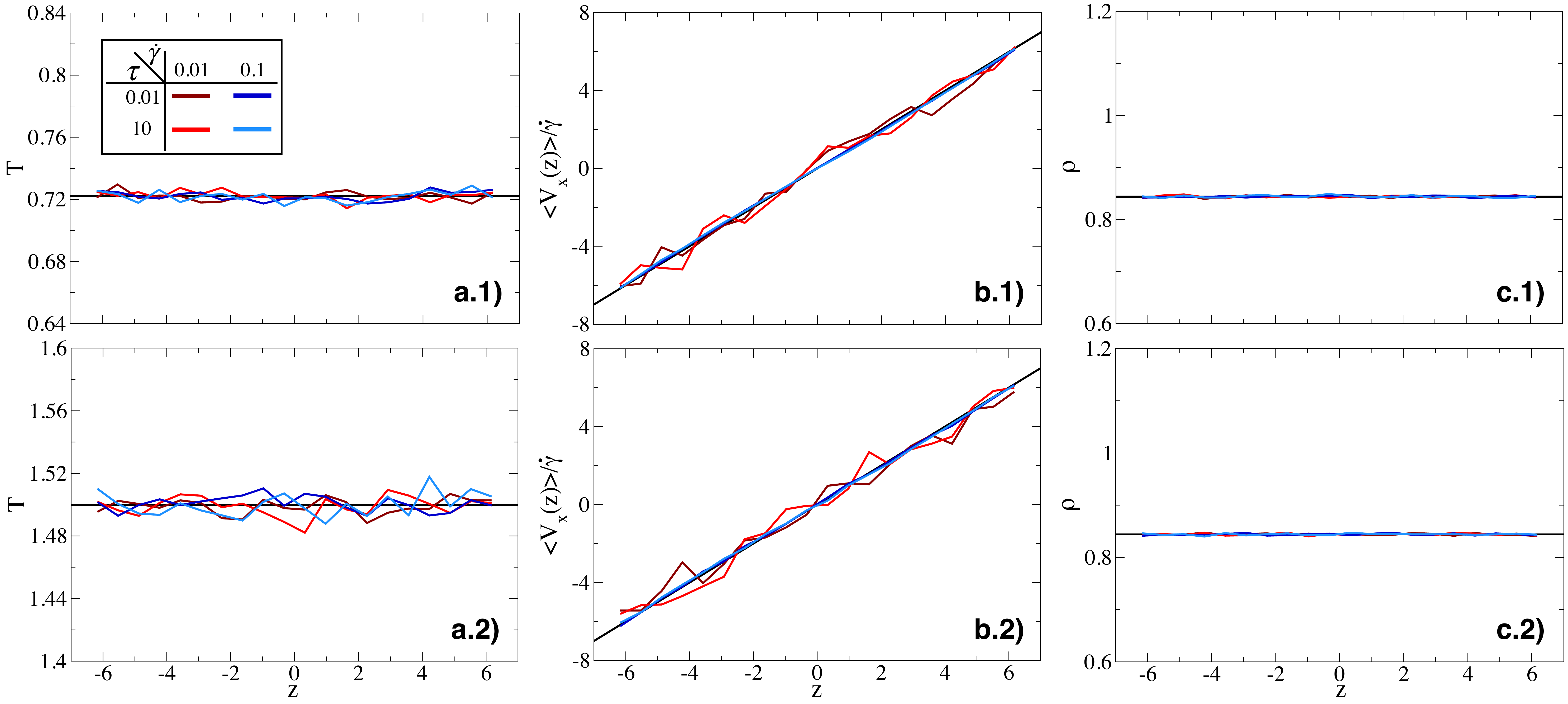}
\caption{(a) Temperature, (b) velocity and (c) density profiles for (a.1, b.1 and c.1) $\dot\gamma=0.01$ and (a.2, b.2 and c.2) $\dot\gamma=0.1$ computed using the BDP thermostat for (top) $T=0.722$ and (bottom) $T=1.5$. The black lines are the reference curves.}
\label{fig:Prof_BUSSI}
\end{figure*}

\subsection{Steady shear with DPD thermostat}

Figures~\ref{fig:Prof_DPD_T0.772_0.01} shows the numerical results under shear flow for systems thermostated with DPD at $T = 0.722$ for several choices of the thermostat parameters. For $\dot\gamma=0.01$ (Fig~\ref{fig:Prof_DPD_T0.772_0.01}.(a-b)) the temperature profile is always constant, regardless of the specific values of the parameters. However, the other two observables show that, once again, a flat $T(z)$ does not guarantee a linear velocity profile. In fact, we observe that the velocity profile is almost but not quite linear only for the lowest values of $\xi$ and $r_{c}$. Moreover, even though $T(z)$ is constant for all possible parameter combinations, its fluctuations increase at small $r_c$. To make things worse, the density profile often exhibits unrealistic inhomogeneities, as with the original Langevin dynamics. Upon increasing $s$ the homogeneity of the velocity and density profiles improves. 

When we increase the shear rate to $\dot\gamma=0.1$ (Fig~\ref{fig:Prof_DPD_T0.772_0.01}.(c-d)), the situation worsens. Indeed, the temperature profile remains constant only at high values of $\xi$ and $r_{c}$, \textit{i.e.} when the thermostat is more coupled (or, equivalently, when the viscosity is higher). In contrast with the Langevin thermostat, here the variation of $\xi$ has a less dramatic impact, as the behaviour of the velocity, temperature and density profiles is fairly insensitive to changes in $\xi$.

The data at $T = 1.5$ display qualitatively similar trends and are not shown here.

\subsection{Steady shear with the modified DPD}

We have seen that increasing the dissipation through a change of the DPD parameters helps in controlling the temperature but negatively affects the velocity profile. The origin of this issue, which has been observed and described in recent work~\cite{chatterjee2007modification,leimkuhler2016pairwise}, can be ascribed to a wrong handling of the periodic boundary conditions along the gradient direction $z$. Indeed, when the $z$-component of a particle $i$ position is close to $\pm L/2$, some of its neighbours are bound to be on the other side of the simulation box boundary. All these will have a streaming velocity that is opposite to the one of $i$, breaking the assumption of translational invariance and of homogeneity. Indeed, as noted in Refs.~\cite{morriss2013statistical} and~\cite{leimkuhler2016pairwise}, particles should not be able to sense when they cross box boundaries, so as to avoid surface effects. The problem can be mitigated by turning off the thermostat for all those pairs of particles that are on different sides of the boundary~\cite{barrat2000fluctuation,varnik2006structural,chatterjee2007modification}. However, there exists a less invasive approach that gets rid of the issue altogether. The idea is to correct the relative velocity between boundary-separated pairs of particles by adding a term $- \textrm{round}(z_{ij}/L) \dot\gamma L$ that takes into account the velocity difference between the top and the bottom of the box~\cite{leimkuhler2016pairwise}.  Figure~\ref{fig:Prof_DPD_MOD} demonstrates that this approach results in a much better control over the temperature, velocity and density profiles.

\subsection{Steady shear with BDP thermostat}

While the original implementation of the BDP thermostat works flawlessly in equilibrium~\cite{bussi2007canonical}, it needs to be adapted to simulations performed under shear flow. First of all, the instantaneous kinetic energy $K$ should not take into account the velocity component along the flow direction. In addition, during the thermalization step, in a fashion similar to what has been done for the peculiar version of the Langevin thermostat, the shear velocity $u_x(z) = \dot\gamma v_z$ is subtracted from the $v_x$ of each particle $i$, the thermostat is applied and then the flow velocity is added back to the new $v_x$~\cite{zausch2009dynamics}. With this change, the BDP thermostat becomes a PBT.

Figure~\ref{fig:Prof_BUSSI} shows the performance of the shear-flow version of the BDP thermostat for the two investigated temperatures and for two different values of $\tau$. We have already seen that $\tau$ has no effect on the dynamics of the system in equilibrium (see Fig.~\ref{fig:MSD}). Similarly, here we see that the BDP thermostat always works well for the combinations of $T$ and $\dot\gamma$ considered here, regardless of the specific value of $\tau$. However, we recall that the BDP thermostat we consider has a global nature and hence does not reproduce hydrodynamics. This thermostat is thus only recommended to generate initial configurations or reference data.

\section{Conclusions}
\label{sec:conclusions}

\begin{table}[]
\centering
(a)\\
\begin{tabular}{|c|c|c|c|c|c|}
\hline
\multicolumn{2}{|c|}{Parameters} & \multicolumn{4}{|c|}{Observables}\\ \hline
$\xi$ & $\dot\gamma$ & $T=0.722$ & $T=1.5$ & \multicolumn{1}{l|}{$v_{x}\left(z\right)$} & $\rho$ \\ \hline
1 & 0.01    & \cmark       & \cmark     & \xmark   & \cmark    \\ \hline
1 & 0.1     & \xmark       & \xmark     & \xmark   & \cmark    \\ \hline
$10^2$ & 0.01  & \cmark       & \cmark     & \xmark & \xmark     \\ \hline
$10^2$ & 0.1 & \cmark       & \cmark     & \xmark & \xmark     \\ \hline
\end{tabular}\\[0.5cm]

(b)\\
\begin{tabular}{|c|c|c|c|c|c|}
\hline
\multicolumn{2}{|c|}{Parameters} & \multicolumn{4}{|c|}{Observables}\\ \hline
$\xi$ & $\dot\gamma$ & $T=0.722$ & $T=1.5$ & \multicolumn{1}{l|}{$v_{x}\left(z\right)$} & $\rho$ \\ \hline
1 & 0.01       & \cmark       & \cmark     & \cmark & \cmark    \\ \hline
1 & 0.1      & \xmark        & \xmark      & \cmark & \cmark    \\ \hline
$10^2$ & 0.01       & \cmark       & \cmark     & \cmark & \cmark    \\ \hline
$10^2$ & 0.1      & \cmark        & \cmark      & \cmark & \cmark    \\ \hline
\end{tabular}
\caption{Assessment of the Langevin dynamics (a) with and (b) without the subtraction of the peculiar velocity for different values of $\xi$ and $\dot\gamma$. The \cmark$\,$ and \xmark$\,$ symbols tell whether the combination of parameters reproduce the correct behaviour for that specific observable or not, respectively.}
\label{table:Langevin}
\end{table}

\begin{table}[]
\centering
\begin{tabular}{|c|c|c|c|c|c|c|c|}
\hline
\multicolumn{4}{|c|}{Parameters} & \multicolumn{4}{|c|}{Observables}\\ \hline
$s$ & $r_c$ & $\xi$ & $\dot\gamma$ & $T=0.722$ & $T=1.5$ & \multicolumn{1}{l|}{$v_{x}\left(z\right)$} & $\rho$ \\ \hline
0.50 & 1.12 & 25 & 0.01 & \cmark       & \cmark     & \cmark                                         & \cmark    \\ \hline
0.50 & 1.12 & 25 & 0.1  & \xmark        & \xmark      & \cmark                                         & \cmark    \\ \hline
1.0 & 1.12 & 25 & 0.01  & \cmark       & \cmark     & \cmark                                         & \cmark     \\ \hline
1.0 & 1.12 & 25 & 0.1   & \xmark       & \xmark     & \cmark                                         & \cmark     \\ \hline
0.50 & 1.88 & 25 & 0.01 & \cmark       & \cmark     & \xmark                                        & \cmark    \\ \hline
0.50 & 1.88 & 25 & 0.1  & \cmark        & \cmark      & \xmark                                        & \xmark    \\ \hline
1.0 & 1.88 & 25 & 0.01  & \cmark       & \cmark     & \xmark                                        & \cmark    \\ \hline
1.0 & 1.88 & 25 & 0.1   & \xmark       & \cmark     & \xmark                                        & \xmark    \\ \hline
\end{tabular}
\caption{Assessment of DPD thermostat (without the modification) considering the set parameter $\left(s,\xi,r_{c}\right)$ values. We only consider the most common values for $\xi$ and $s$. The \cmark$\,$ and \xmark$\,$ symbols tell whether the combination of parameters reproduce the correct behaviour for that specific observable or not, respectively.}
\label{table:DPD}
\end{table}

We have shown that, when studying out-of-equilibrium systems, having a constant temperature profile is a necessary but not sufficient condition to ensure the correctness of the simulation. Indeed, one has to be sure that also the velocity and density profiles exhibit sound behaviours. Here we have tested several thermostats under different physical and profile conditions. In general, we observe that, under PUT conditions, a fine tuning of the thermostat parameters is necessary to avoid non-physical behaviours. When $\dot{\gamma}$ increases more energy is pumped into the system and hence the thermostat has to be more tightly coupled to the system in order to maintain the desired temperature. If the thermostat acts on the absolute velocities, a too-high friction might cancel the effect of the imposed shear flow on the particles that are far from the box boundaries, which is where the flow velocity acquires the highest values. This, in turn, affect the density profile, which becomes inhomogeneous. 

By contrast, if the thermostat is applied on the peculiar velocities, that is, if we use a PBT, correct velocity and density profiles will be assured. If we need to explore high values of $\dot{\gamma}$ it is recommended to work with a PBT. However, doing so will make it impossible to observe some real physical phenomena such as shear banding. For the Langevin thermostat, PBT conditions can be implemented by removing the peculiar (flow) velocity in the friction term (see Eq.~\eqref{eq:Peculiar}). A similar remark holds true for the Bussi-Donadio-Parrinello thermostat, which turned out to be the most stable and reliable thermostat. However, the BDP thermostat is also the one exhibiting the less realistic dynamics. Tables~\ref{table:Langevin} and~\ref{table:DPD} summarise some of the results reported.

The best choice in terms of stability, realism and computational efficiency is probably the DPD thermostat, modified so as to take into account the relative difference between the streaming velocities at the boundary of the simulation box~\cite{leimkuhler2016pairwise}. However, this thermostat also depends on the largest number of parameters: the exponent $s$, the cut-off $r_c$ and the friction constant $\xi$. According to our results, the common choice of setting $s = 0.5$ or $1$ gives overall good results. One of the main difficulties is to choose an optimal value for $r_{c}$. There is not an \textit{a priori} physical motivation to choose a certain value, and the optimal choice of $r_c$ is in general independent on the other parameters. In general, making a good choice requires a knowledge of the structure of the system under study. For instance, if $r_c$ is much smaller than the average particle-particle distance, which might happen in dilute systems, the thermostat is essentially decoupled from the system and might not be able to dissipate the extra energy. Our results show that the cut-off radius should be between the first maximum and the first minimum of the $g(r)$, and that too-large values might originate non-physical behaviour such as anomalies and inhomogeneities in the velocity and density profiles. A value smaller than but close to the first minimum of the radial distribution function of the system is a good starting point.


Throughout the paper we have not analysed the computational efficiency of the different thermostats, which greatly depends on the computational (\textit{e.g.} serial \textit{vs} parallel codes) and model (\textit{e.g.} short- \textit{vs.} long-ranged potentials) details. However, for the systems studied here the BDP thermostat turned out to be the most efficient thermostat, providing an additional reason to use it to produce exploratory data and to prepare initial configurations for the production runs. Second comes the Langevin thermostat. The DPD thermostat, which features pair interactions, comes last, even though its efficiency can be improved by using \textit{e.g.} Verlet neighbouring lists~\cite{verlet1967computer}. Given the great number of random numbers required by the DPD thermostat, it might be tempting to extract them from uniform instead of Gaussian distributions~\cite{dunweg1991brownian}. However, we advise against it, as doing so yields the wrong temperature profile for some combinations of the thermostat parameters.

\section*{Author contribution statement}
JRF, LR and EZ performed simulations, analysed results and wrote the paper.

\section*{Acknowledgements}
JRF and EZ acknowledge support from ETN-COLLDENSE (H2020-MCSA-ITN-2014, Grant No. 642774). LR, EZ acknowledge support from the ERC Consolidator Grant 681597 MIMIC. We thank N. Gnan for useful discussions.

\bibliography{biblio}

\end{document}